# Effective Young's Modulus of Two-Phase Elastic Composites by Repeated Isostrain/Isostress Constructions and Arithmetic-Geometric Mean

Jiashi Yang (jyang1@unl.edu)
Department of Mechanical and Materials Engineering
University of Nebraska-Lincoln, Lincoln, NE 68588-0526, USA

A relationship is established between the effective Young's modulus of a two-phase elastic composite and a known mathematical mean value. Specifically, the effective Young's modulus of a composite obtained from repeated parallel/serial constructions is equal to the arithmetic-geometric mean of the Young's moduli of the component materials. This result also applies to electric circuits with resistors in repeated parallel/serial connections.

As a necessary mathematical background, we summarize the basics of the arithmetic-geometric mean below [1]. Consider two positive numbers, $x$ and $y$ with $x>y$, from which the following two sequences can be defined:

$$a_0 = x, \quad g_0 = y,$$
$$a_{n+1} = \frac{1}{2}(a_n + g_n), \quad g_{n+1} = \sqrt{a_n g_n}, \tag{1}$$
$$n = 0, 1, 2, \cdots.$$

It is known that $a_n$ and $g_n$ converge to the same limit, the arithmetic-geometric mean of $x$ and $y$, denoted by $M(x,y)$.

Consider a composite of two isotropic elastic materials with Young's modulus $E_A$ and $E_B$ as well as volume fractions $\phi_A$ and $\phi_B$, respectively. Using one-dimensional models for parallel (isostrain or Voigt) and serial (isostress or Reuss) constructions, we have the following effective elastic constants [2], $E_R$ and $E_V$:

$$E_V = \phi_A E_A + \phi_B E_B, \tag{2}$$

$$\frac{1}{E_R} = \phi_A \frac{1}{E_A} + \phi_B \frac{1}{E_B} \quad \left(\text{or} \quad E_R = \frac{E_A E_B}{\phi_A E_B + \phi_B E_A}\right), \tag{3}$$

and

$$E_R \leq E_V, \tag{4}$$

which provide lower and upper bounds for the effective Young's modulus of composites with other constructions [2].

It is natural to wonder what if one continues the parallel and serial constructions repeatedly beginning with $E_R$ and $E_V$. Therefore, we let

$$p_0 = E_V, \quad s_0 = E_R,$$
$$p_{n+1} = \frac{1}{2} p_n + \frac{1}{2} s_n, \quad s_{n+1} = \frac{p_n s_n}{\frac{1}{2} p_n + \frac{1}{2} s_n}, \tag{5}$$
$$n = 0, 1, 2, \cdots,$$

where, since the initial volume fractions have been taken into consideration in $p_0$ and $s_0$ through (2) and (3), the same volume of the two materials are used in subsequent parallel and serial constructions.

(5) defines two sequences which are related to the $a_n$ and $g_n$ in (1) through

$$p_n = a_n, \quad s_n = \frac{g_n^2}{a_n}. \tag{6}$$

Since $a_n$ and $g_n$ have the same limit, then

$$\lim_{n\to\infty} p_n = \lim_{n\to\infty} s_n = M(E_V, E_R). \tag{7}$$

Hence, for a composite so obtained from repeated parallel/serial constructions beginning with $E_R$ and $E_V$, the effective Young's modulus is the arithmetic-geometric mean of $E_V$ and $E_R$.